\documentclass[11pt]{article}

\PassOptionsToPackage{numbers,sort&compress}{natbib}
\usepackage[preprint]{acl}
\usepackage{times}
\usepackage{latexsym}
\usepackage[T1]{fontenc}
\usepackage[utf8]{inputenc}
\usepackage{microtype}
\usepackage{inconsolata}
\usepackage{graphicx}
\usepackage{url}
\usepackage{doi}
\usepackage{array}
\usepackage{amssymb}
\usepackage{amsmath}
\usepackage{amsthm}
\usepackage{booktabs}
\usepackage{tabularx}
\usepackage{caption}
\usepackage{subcaption}
\usepackage[ruled,vlined,linesnumbered]{algorithm2e}
\usepackage{tikz}
\usepackage{pgfplots}
\usetikzlibrary{positioning, arrows.meta, shapes.geometric, fit, backgrounds, calc}
\pgfplotsset{compat=1.18}

\title{LLM-Guided Prompt Evolution for Password Guessing \\
\bigskip \normalsize{{\textcolor{red}{Warning: This paper contains potentially harmful content generated by LLMs.}}}}

\author{
  Vladimir A. Mazin \\
  AXXX 
  \And
  Mikhail A. Zorin \\
  MTUCI \\ 
  \And
  Dmitrii S. Korzh \\
  AXXX
  \AND
  Elvir Z. Karimov \\
  Applied AI Center
  \And
  Dmitrii A. Bolokhov \\
  MTUCI \\ 
  \And
  Oleg Y. Rogov \\
  AXXX \\ Applied AI Center
}

\begin{document}
\maketitle

\begin{abstract}
Passwords still remain a dominant authentication method, yet their security is routinely subverted by predictable user choices and large-scale credential leaks. Automated password guessing is a key tool for stress-testing password policies and modeling attacker behavior. This paper applies LLM-driven evolutionary computation to automatically optimize  prompts for the LLM password guessing framework. Using OpenEvolve, an open-source system combining MAP-Elites quality-diversity search with an island population model  we evolve prompts that maximize cracking rate on a RockYou-derived test set. We evaluate three configurations: a local setup with Qwen3 8B, a single compact cloud model Gemini-2.5 Flash, and a two-model ensemble of frontier LLMs. The approach raises the cracking rates from 2.02\% to 8.48\%. Character distribution analysis further confirms how evolved prompts produce statistically more realistic passwords. Automated prompt evolution is a low-barrier yet effective way to strengthen LLM-based password auditing and underlining how attack pipelines show tendency via automated improvements.
\end{abstract}

\section{Introduction}
\label{sec:intro}

Although increasingly complex methods of identity verification are emerging, passwords remain the primary means of authentication for the vast majority of internet users and corporate systems~\cite{bonneau2012quest}. Large-scale data leaks involving access credentials are becoming increasingly common: the RockYou hack alone exposed 32 million passwords in plain text, and subsequent data leaks such as Collection\#1 exposed over a billion unique login credentials~\cite{hunt2019collection}. These data leaks directly facilitate offline attacks using dictionaries and rule-based cracking attacks, as many users choose predictable passwords and reuse credentials for different services.

In this context, automated password guessing tools serve a dual purpose: they help security experts verify password policies and identify weak credentials before attackers do, while also capturing the increasing sophistication of real-world attackers---a tension familiar from surveys of LLMs in cybersecurity~\cite{zhang2025llmcybersecurity} and from holistic treatments of model misuse and safety~\cite{shi2024llmsafety}. Classic approaches -- dictionary attacks, Markov models~\cite{narayanan2005markov} and probabilistic context-free grammars (PCFG)~\cite{weir2009pcfg} -- are effective, but cannot be generalised beyond their training distribution. Neural network methods, including RNN-based guessers~\cite{melicher2016fast} and GANs~\cite{hitaj2019passgan}, capture richer structural patterns, but their fixed architectures limit their adaptability to new contexts.

The introduction of PassLLM~\cite{zou2025password} marked a qualitative shift: large language models (LLMs) fine-tuned on leaked password corpora via LoRA outperform all previous methods in both mass (trawling) and targeted guessing scenarios. However, PassLLM's behaviour during inference is entirely controlled by a short text prompt, whereas that study reports that ``the content of the prompt has minimal impact on the results of the guessing task''---a claim we directly challenge in this work.

The manual creation of prompts for security tasks is both time-consuming and domain-sensitive: the optimal prompt for a password-generating LLM is not obvious and is highly sensitive to wording, structural cues, and contextual constraints. We therefore ask the question: Can \emph{automated, evolutionary} optimization of the prompt reliably improve PassLLM beyond what a manually crafted prompt achieves?

We answer this question in the affirmative by integrating PassLLM with OpenEvolve~\cite{openevolve2025}, an open-source LLM-driven evolutionary framework that combines MAP-Elites~\cite{mouret2015illuminating} quality-diversity search with a multi-island population model. 

\paragraph{Research questions.}
\label{sec:rqs}
\noindent We structure our investigation as follows.
\begin{description}
  \item[\textbf{RQ1} (Prompt Sensitivity)] Does the content of the inference prompt
    significantly influence the cracking rate of PassLLM, contradicting the claim of~\cite{zou2025password} that prompt content has minimal impact?

  \item[\textbf{RQ2} (Effectiveness)] Can LLM-driven evolutionary optimization
    reliably improve PassLLM's cracking rate beyond a manually crafted baseline
    prompt within a fixed budget of 100 evolutionary iterations?

  \item[\textbf{RQ3} (Resource Efficiency)] Does a locally constrained
    configuration (single 24\,GB GPU, 8B-parameter model) achieve competitive
    cracking rates compared to cloud-based frontier-model ensembles?

  \item[\textbf{RQ4} (Distribution Realism)] Do evolved prompts produce password
    populations whose character-frequency distributions are statistically closer
    to real-world corpora, as measured by the per-symbol F-score AUC metric?
\end{description}





The rest of the paper is organized as follows.
Section~\ref{sec:related} provides an overview of related work, including LLM safety and misuse (Section~\ref{sec:llm_safety}).
Section~\ref{sec:threat} introduces the threat model.
Section~\ref{sec:model} describes the integrated pipeline.
Section~\ref{sec:experimental_setup} describes the experimental setup in detail.
Section~\ref{sec:results} presents the results, explicitly mapping each finding to the research questions in Section~\ref{sec:rqs}.
Section~\ref{sec:discussion} contains an in-depth analysis and discussion.
Section~\ref{sec:limitations} states limitations and ethical considerations.
Section~\ref{sec:conclusion} contains the conclusion.

\section{Related Work}
\label{sec:related}

\subsection{Password guessing: classical methods}
Early automated password cracking methods were based on offline dictionary attacks and rule-based engines such as John the Ripper~\cite{openwall2025jtr} and Hashcat~\cite{hashcat2025}, which are still widely used due to their efficiency on commercially available hardware. Probabilistic context-free grammars (PCFG) were introduced~\cite{weir2009pcfg} to model the structural decomposition of passwords, thereby significantly improving coverage over pure dictionary methods. Markov chains applied to time-space trade-offs help achieving fast, high-quality guessing~\cite{narayanan2005markov}. In subsequent work, PCFG was extended by incorporating semantic structures and signals about user behaviour~\cite{huang2023se}.

\subsection{Neural and generative approaches}
Recurrent neural networks (RNNs) trained on leaked corpora can model the sequential structure of passwords and offer competitive coverage with low memory requirements~\cite{melicher2016fast}. PassGAN~\cite{hitaj2019passgan} applies generative adversarial networks to password synthesis; although training GANs is difficult to stabilise~\cite{saxena2020gans}, PassGAN demonstrated that adversarial generation can discover passwords not included in the training dictionary. More recent Transformer-based approaches such as PassBERT~\cite{ming2023passbert} use bi-directional transformers with a pre-training/fine-tuning paradigm to improve real-world password guessing attacks, including conditional, targeted, and adaptive rule-based guessing.

\subsection{LLM-based password guessing}
PassLLM~\cite{zou2025password} applies LoRA-based fine-tuning to models from the Llama, Mistral, and Qwen families on the RockYou dataset. PassLLM outperforms all previous methods in terms of trawling, PII-based (personally identifiable information), and password reuse attack scenarios. However, that work treats prompt selection as a minor implementation detail and claims that sensitivity to prompt content is minimal---a premise we experimentally refute.

\subsection{LLM safety and misuse}
\label{sec:llm_safety}
Integration of LLMs into authentication-related research sits at the intersection of defensive evaluation and dual-capability tooling~\cite{shi2024llmsafety}. Systematic reviews of LLM applications in cybersecurity catalog both protective uses and assisted offensive workflows~\cite{zhang2025llmcybersecurity}. Fine-tuning on security-oriented instruction data can improve task accuracy while sharply degrading robustness to prompt-based attacks~\cite{elzemity2025cyberllm}, and post-alignment adaptation does not reliably preserve refusal behaviour~\cite{qi2024finetuning}. Beyond weight updates, natural-language context is itself a powerful control surface: indirect prompt injection shows how untrusted text can override intended instructions in integrated systems~\cite{greshake2023indirect}, and narrow in-context demonstrations can induce broadly misaligned outputs on unrelated queries~\cite{afonin2026emergent}. Mechanistic interventions are similarly ambiguous---activation steering, though often motivated as interpretable control, can systematically weaken safeguards~\cite{korznikov2026rogue}. Gradient-free search over short text strings can discover transferable adversarial suffixes for aligned chat models~\cite{zou2023universal}, foreshadowing automated prompt optimisation for specialised generative tasks such as password guessing. Frameworks for evaluating frontier models for extreme risks explicitly include offensive cyber capabilities among the hazards warranting structured assessment~\cite{shevlane2023extremerisks}.

\subsection{Evolutionary and LLM-driven optimization}
Evolutionary computation has a long history in combinatorial optimization, and its combination with LLMs is a new and rapidly evolving paradigm. The OpenEvolve framework~\cite{openevolve2025} implements MAP-Elites~\cite{mouret2015illuminating} quality-diversity search with LLM-guided mutations and achieves remarkable results in code optimization and mathematical discovery. Its architecture is formally analyzed
in~\cite{brown2025even}. LoongFlow~\cite{wan2026loongflow} extends this paradigm with a cognitive plan-execute-summarize loop and demonstrates improved convergence. In the field of NLP,  automatic prompt optimization has been explored in the context of the Automatic Prompt Engineer (APE) framework~\cite{zhou2022ape} and gradient-free methods such as PromptBreeder~\cite{fernando2023promptbreeder}. The proposed pipeline is the first to apply LLM-driven quality-diversity evolution specifically to
the field of cybersecurity in password guessing.

\section{Threat Model}
\label{sec:threat}
We use the following setting in our experimental setup.

\subsection{Attacker's Goals and Scenario}
We consider a trawling attack (mass attack) in which an attacker targets many accounts in order to compromise as many accounts as possible within a fixed budget of $B = 20,000$ attempts per account. This is the standard threat scenario evaluated by PassLLM~\cite{zou2025password} and strength metrics based on guesses~\cite{dell2010password}.

\subsection{Attacker's capabilities}
The attacker works offline with a hacked, hashed password database. Guesses are made locally under GPU time and memory constraints. The attacker has access to public datasets (RockYou, LinkedIn, Collection\#1) for training and white-box control over PassLLM, including prompts. For prompt engineering, the attacker can use cloud APIs (e.g., OpenRouter) or local open-weight models (e.g., Ollama). We assume a modest budget -- an NVIDIA GPU with 24 GB VRAM and up to 100 evolution iterations -- reflecting a cost-conscious adversary.

\subsection{Attacker's knowledge}
The attacker knows the architecture and weights of PassLLM (including LoRA and \texttt{Qwen2.5-0.5B-Instruct}) and has a small set of target passwords (e.g., previously compromised hashes). Knowledge of users' personal data, hash functions, or website policies is not necessary, as they are irrelevant to offline trawling. Protective measures such as password policies, rate limiting, and hashing methods are beyond the attacker's control.

\subsection{Implications for defence}
This model highlights two defensive insights. First: strict password policies, e.g., blocking leaked passwords worsen the attacker's fitness signal and slow down development~\cite{florencio2014password}. Second: defenders may not rely on static tools, as the proposed method improves cracking rates in just 1--10 iterations, in line with broader evidence that model behaviour can shift rapidly under optimised context~\cite{zou2023universal,afonin2026emergent}. Adaptive rate limiting and anomaly detection remain crucial.

\section{System Description}
\label{sec:model}

\begin{figure*}
\centering
\includegraphics[width=0.8\linewidth]{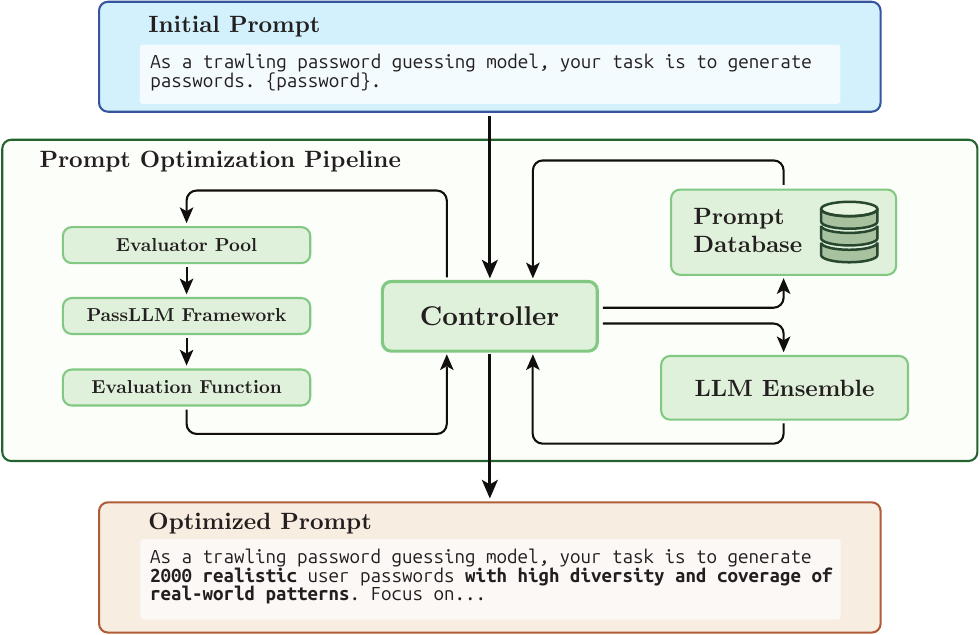}
\caption{The proposed OpenEvolve-based pipeline. An initial prompt is fed to the evolutionary controller, which iteratively samples parent prompts from the MAP-Elites archive, requests improved variants from the LLM ensemble, runs LLM and updates the archive based on the cracked-rate fitness signal.}
\label{fig:openevolve_passllm_flowchart}
\end{figure*}

This section describes the integrated PassLLM and OpenEvolve pipeline and formalises the evolutionary
process for optimising prompts. An overview of the pipeline is shown in Fig.~\ref{fig:openevolve_passllm_flowchart}.

\subsection{The LLM guessing model}
\label{sec:passllm}

We focus on the trawling variant of PassLLM~\cite{zou2025password}, which is based on
\texttt{Qwen2.5-0.5B-Instruct} and was fine-tuned via LoRA on the publicly available RockYou split of the PassLLM. Inference employs a breadth-first search over the model's token lattice to
efficiently enumerate the most probable strings. The model's generation behaviour is
completely determined by a single command input $\mathbf{p}$; this is the optimization target
of the proposed evolutionary procedure.

\subsection{Target function}
\label{sec:fitness}

Given an input $\mathbf{p}$, LLM generates a multiset of password candidates
$\mathcal{G}(\mathbf{p}) = \{g_1, \ldots, g_B\}$ where B is under $2\times10^4$. The cracked rate (CR) is selected as the primary fitness function:
\begin{equation}
  \mathrm{CR}(\mathbf{p}) = \frac{|\mathcal{G}(\mathbf{p}) \cap \mathcal{T}|}{|\mathcal{T}|},
  \label{eq:cr}
\end{equation}
where $\mathcal{T}$ denotes a fixed holdout test set of passwords from the RockYou database.
$|\mathcal{G}(\mathbf{p}) \cap \mathcal{T}|$ counts exact matches of character strings, taking case sensitivity into account.
This metric directly quantifies the attacker's success rate under a model with limited attempts.

\subsection{Complementary metrics}
\label{sec:fscore}

CR alone does not provide any information about why a prompt is effective. To gain deeper insight into the distribution properties of the generated passwords at the character level, we introduce a supplementary
F-score metric per symbol. For each character $c$ in the printable ASCII alphabet,
$f_c^{\mathrm{gen}}$ and $f_c^{\mathrm{real}}$ denote its normalised frequency in the generated
and real password sets, respectively. We define a binary `hit' for the character $c$ at a threshold value
$\tau$ as $\mathbf{1}[f_c^{\mathrm{gen}} \geq \tau \cdot f_c^{\mathrm{real}}]$ and calculate the
macro-averaged F-score across all characters:
\begin{equation}
  \mathrm{F}_\tau(\mathbf{p}) = \frac{2 \cdot \mathrm{Prec}_\tau(\mathbf{p}) \cdot \mathrm{Rec}_\tau(\mathbf{p})}
    {\mathrm{Prec}_\tau(\mathbf{p}) + \mathrm{Rec}_\tau(\mathbf{p})},
  \label{eq:fscore}
\end{equation}
where precision and recall are computed over the set of characters above the frequency threshold $\tau$.
By sweeping $\tau \in [0, 1]$ we obtain an F-score curve analogous to a precision-recall curve,
and we summarize it with the area under the curve.

\subsection{Evolutionary framework: OpenEvolve}
\label{sec:openevolve}

Prompt evolution is carried out with OpenEvolve~\cite{openevolve2025}, an open-source framework that combines two complementary mechanisms. OpenEvolve uses a MAP-Elites~\cite{mouret2015illuminating} grid ($10^2$ cells per island) to store the prompts with the highest fitness for combinations of three features: complexity (number of tokens), diversity (Levenshtein distance to the reference), and prompt length. This maintains quality diversity across the entire feature space. Three independent islands each run MAP-Elites and exchange 10\% of the elite prompts every 10 iterations in a ring topology. Instead of random character changes, an LLM performs semantic mutation: starting from parent prompts and a meta-prompt describing the goal, it generates variants. This allows for both fine-tuning and structural rewrites, helping to avoid local optima.
The selection of parents at each generation is based on a three-part mixture:
\begin{itemize}
  \item \textbf{Elite selection} (ratio 0.1) -  highest fitness prompts from the archive;
  \item \textbf{Exploration} (ratio 0.2) -  random samples from the current island population;
  \item \textbf{Exploitation} (ratio 0.7) - samples from archive cells weighted by fitness.
\end{itemize}
The heavy exploitation bias is appropriate for proposed setting, where evaluation cost (PassLLM inference)
is high and we wish to maximize fitness efficiently.
Algorithm~\ref{alg:evo_pipeline} describes the evolutionary prompt optimization pipeline.

\begin{algorithm}[!tbh]
\DontPrintSemicolon
\SetAlgoLined
\SetKwInOut{Require}{Require}
\SetKwInOut{Ensure}{Ensure}
\SetKwFor{ParFor}{for}{do (parallel)}{end}
\SetKwFunction{Select}{Select}
\SetKwFunction{LLMMutate}{LLMMutate}
\SetKwFunction{CR}{CR}
\SetKwFunction{FeatureVec}{FeatureVec}
\SetKwFunction{UpdateArchive}{UpdateArchive}

\Require{Initial prompt $\mathbf{p}_0$, test set $\mathcal{T}$, budget $B$, max iterations $T$,
         islands $K$, migration interval $\Delta$, migration rate $\mu$, LLM ensemble $\mathcal{M}$}
\Ensure{Best prompt $\mathbf{p}^*$ maximising $\mathrm{CR}(\mathbf{p})$}

\BlankLine
Initialize MAP-Elites archive $\mathcal{A}_k \leftarrow \emptyset$ \textbf{for} $k = 1, \ldots, K$\;
$s_0 \leftarrow \CR(\mathbf{p}_0)$ \tcp*[r]{$s_0 = 2.02\%$ in the proposed setting}
Insert $\mathbf{p}_0$ into $\mathcal{A}_k$ for all $k$\;

\BlankLine
\For{$t \leftarrow 1$ \KwTo $T$}{
    \ParFor{each island $k = 1, \ldots, K$}{
        $\mathbf{p}_{\mathrm{par}} \leftarrow \Select(\mathcal{A}_k)$
            \tcp*[r]{elite, explore, exploit, mixture}
        $\mathbf{p}_{\mathrm{new}} \leftarrow \LLMMutate(\mathbf{p}_{\mathrm{par}},\, \mathcal{M})$\;
        $s_{\mathrm{new}} \leftarrow \CR(\mathbf{p}_{\mathrm{new}})$
            \tcp*[r]{run PassLLM, count matches vs.\ $\mathcal{T}$}
        $\phi \leftarrow \FeatureVec(\mathbf{p}_{\mathrm{new}})$\;
        $\UpdateArchive(\mathcal{A}_k,\, \mathbf{p}_{\mathrm{new}},\, s_{\mathrm{new}},\, \phi)$\;
    }
    \If{$t \bmod \Delta = 0$}{
        Migrate elites between islands:\quad
        $\mathcal{A}_k \overset{\mu}{\longleftrightarrow} \mathcal{A}_{(k+1)\bmod K}$\;
    }
}

\BlankLine
$\mathbf{p}^{*} \leftarrow \displaystyle\arg\max_{\mathbf{p}\,\in\,\bigcup_k \mathcal{A}_k}
    \mathrm{CR}(\mathbf{p})$\;
\Return{$\mathbf{p}^{*}$}\;

\caption{Evolutionary Prompt Optimization for PassLLM}
\label{alg:evo_pipeline}
\end{algorithm}

\subsection{Models for prompt generation}

Various LLMs served as mutation operators. All models were accessed either locally via
Ollama or via a cloud service with a fixed OpenAI-compatible API.
The generation parameters were set: \texttt{temperature} $= 0.4$ and \texttt{max\_tokens} $= 16000$ for all
experiments to ensure comparability. Three configurations were evaluated:
\begin{enumerate}
  \item \textbf{Local (Ollama):} \texttt{qwen3:8b} for both generation and evaluation.
    Single 24 GB GPU. 100 evolutionary iterations.
  \item \textbf{Cloud, single model:} \texttt{gemini-2.5-flash-lite-preview-09-2025} for generation;
    \texttt{qwen3-max} for evaluation. 100 iterations.
  \item \textbf{Cloud, ensemble:} Weighted mixture of \texttt{qwen3-235b-a22b-2507} (weight~0.5)
    and \texttt{gemini-3-flash-preview} (weight~0.5) for generation; \texttt{qwen3-max} for
    evaluation. 100 iterations.
\end{enumerate}
Separating the generation and evaluation of LLMs reduces systematic biases caused by a single model architecture
and provides a more objective fitness signal.

The complete hyperparameter settings are listed in Table~\ref{tab:openevolve_hyperparams}.

\section{Experimental Setup}
\label{sec:experimental_setup}

\subsection{Dataset and evaluation protocol}

All experiments use the publicly available RockYou password data set in the train/test split,
provided by the PassLLM authors~\cite{zou2025password}. The test set contains passwords
that were not seen during the LoRA fine-tuning of PassLLM. The cracked rate (Equation~\ref{eq:cr}) is the primary
evaluation metric; the F-score per symbol (Equation~\ref{eq:fscore}) is calculated for the best prompt
of each experiment on the same test set. For each evolution run, PassLLM generates up to $B = 20000$ candidate passwords per evaluation,
with beam search batch sizes ranging from 256 to 512 to fit the 24 GB vRAM budget. All experiments are conducted with a single nVidia A100 GPU.


PassLLM~\cite{zou2025password} was   selected as the baseline for the experiments with its \emph{original, unmodified} prompt:
\begin{center}
\texttt{``As a trawling password guessing model, your task is to generate passwords. \{password\}.''}
\end {center}


Expermental parameters are specified in Open Evolve YAML configuration file and will be included in the GitHub repository alongside the code for the reproducibility. The most important parameters
are presented in the Table~\ref{tab:openevolve_hyperparams}.

\begin{figure*}[t]
\centering
\resizebox{0.75\textwidth}{!}{%
\begin{tikzpicture}[
  island/.style={draw=gray!60, rounded corners=3pt, thick},
  migr/.style={->, thick, gray!60},
  cell/.style={fill=teal!55, draw=teal!70, thin}
]
\def\islandW{2.2}
\def\islandH{2.2}
\def\gridStep{0.55}

\node[anchor=east, font=\scriptsize\bfseries, inner xsep=2pt, rotate=90]
  at (0, 3.1) {prompt length};
\node[anchor=south, font=\scriptsize\bfseries] at (4.3, 3.6) {complexity};

\begin{scope}[xshift=0.6cm, yshift=0.95cm]
  \draw[island] (0, 0) rectangle (\islandW, \islandH);
  \draw[gray!50, thin] (0, 0) grid[step=\gridStep] (\islandW, \islandH);
  \fill[cell] (0.55, 0.55) rectangle (1.1, 1.1);
  \fill[cell] (1.1, 1.1) rectangle (1.65, 1.65);
  \fill[cell] (1.65, 0.55) rectangle (2.2, 1.1);
  \node[font=\small] at (1.1, -0.2) {Island 0};
\end{scope}
\begin{scope}[xshift=3.2cm, yshift=0.95cm]
  \draw[island] (0, 0) rectangle (\islandW, \islandH);
  \draw[gray!50, thin] (0, 0) grid[step=\gridStep] (\islandW, \islandH);
  \fill[cell] (1.1, 1.1) rectangle (1.65, 1.65);
  \fill[cell] (1.65, 0.55) rectangle (2.2, 1.1);
  \node[font=\small] at (1.1, -0.2) {Island 1};
\end{scope}
\begin{scope}[xshift=5.8cm, yshift=0.95cm]
  \draw[island] (0, 0) rectangle (\islandW, \islandH);
  \draw[gray!50, thin] (0, 0) grid[step=\gridStep] (\islandW, \islandH);
  \fill[cell] (1.1, 1.1) rectangle (1.65, 1.65);
  \fill[cell] (0.55, 1.65) rectangle (1.1, 2.2);
  \fill[cell] (1.65, 0.55) rectangle (2.2, 1.1);
  \node[font=\small] at (1.1, -0.2) {Island 2};
\end{scope}

\draw[migr] (2.82, 2.05) -- (3.2, 2.05);
\draw[migr] (5.42, 2.05) -- (5.8, 2.05);

\node[draw=gray!40, fill=gray!8, rounded corners=3pt, align=center,
  font=\footnotesize, text width=0.7\textwidth] at (4, 5) {
  \textbf{MAP-Elites:} each cell stores one best prompt (elite).
  Axes: \texttt{prompt length}, \texttt{complexity} (bins 0--9).
  Migrations occur periodically.
};
\draw[->, thick, gray!60] (6.9, 3.15) |- (3.1, 3.4) -| (1.7, 3.15);
\node[font=\small, fill=white, inner sep=2pt] at (4.3, 3.4) {cycle};
\draw[gray!50, thin] (3, 0.42) rectangle (3.3, 0.22);
\node[anchor=west, font=\scriptsize, inner xsep=4pt] at (3.25, 0.30) {empty};
\fill[cell] (4.55, 0.42) rectangle (4.85, 0.22);
\node[anchor=west, font=\scriptsize, inner xsep=4pt] at (4.8,0.33) {elite};
\end{tikzpicture}%
}
\caption{MAP-Elites archive and island model. Three islands each maintain an independent
  $\texttt{prompt\_length} \times \texttt{complexity}$ grid. Filled green cells indicate occupied elites. Ring-topology migration occurs at fixed intervals.}
\label{fig:map_elites_islands}
\end{figure*}
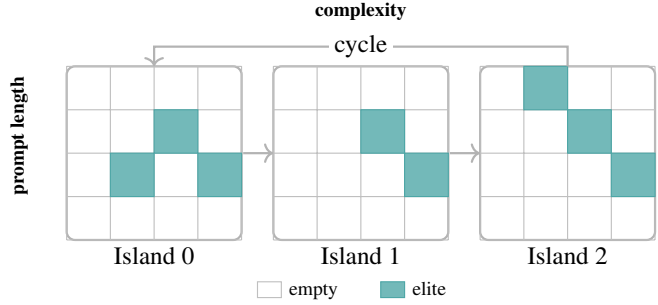

\begin{table*}[t]
  \centering
  \caption{OpenEvolve hyperparameter settings. CR: best cracked rate achieved.}
  \label{tab:openevolve_hyperparams}
  \footnotesize
  \begin{tabularx}{\textwidth}{@{}p{3.4cm}p{2.2cm}p{2.2cm}X@{}}
  \toprule
  \textbf{Parameter} & \textbf{Exp.~1 (Ollama)} & \textbf{Exp.~2 (Gemini)} &
    \textbf{Exp.~3 (Ensemble)} \\
  \midrule
  LLM (generation) & qwen3:8b & gemini-2.5-flash-lite & qwen3-235b (0.5), gemini-3-flash (0.5) \\
  Evaluator LLM & qwen3:8b & qwen3-max & qwen3-max \\
  Temperature & 0.4 & 0.4 & 0.4 \\
  Max tokens & 16\,000 & 16\,000 & 16\,000 \\
  Population size & 100 & 1000 & 1000 \\
  Archive size & 20 & 100 & 100 \\
  Islands ($K$) & 3 & 3 & 3 \\
  Migration interval $\Delta$ & 10 & 10 & 10 \\
  Migration rate $\mu$ & 0.1 & 0.1 & 0.1 \\
  Feature dimensions & diversity, complexity & diversity, complexity &
    prompt\_length, complexity \\
  Feature bins & 10 & 10 & 10 \\
  Elite / explor.\ / exploit.\ & 0.1 / 0.2 / 0.7 & 0.1 / 0.2 / 0.7 & 0.1 / 0.2 / 0.7 \\
  Random seed & 42 & 42 & 42 \\
  Max iterations $T$ & 100 & 100 & 100 \\
  Best CR & 8.28\% & 7.58\% & \textbf{8.48\%} \\
  \bottomrule
  \end{tabularx}
\end{table*}

\section{Results}
\label{sec:results}


\noindent\textbf{To answer RQ1 (prompt sensitivity),} we compare the original PassLLM instruction against evolved prompts under identical inference and evaluation protocols. The baseline cracked rate remains at 2.02\% on the hold-out set, whereas all three evolved configurations achieve substantially higher peak rates (Table~\ref{tab:stats_summary}), demonstrating that prompt content strongly conditions guessing performance.

\noindent\textbf{To answer RQ2 (effectiveness of evolution),} we run OpenEvolve for up to 100 iterations per configuration and record the best cracked rate (CR) in each run. Table~\ref{tab:stats_summary} shows improvements of $3.75\times$--$4.2\times$ over the baseline; Fig.~\ref{fig:evolution_first45} and Fig.~\ref{fig:evolution_full} show that large gains can appear within the first iterations, before the archive stabilises.

\noindent\textbf{To answer RQ3 (resource efficiency),} we contrast the local single-GPU setup (Experiment~1, \texttt{qwen3:8b}) with cloud frontier ensembles (Experiment~3). The local configuration reaches 8.28\% best CR versus 8.48\% for the ensemble---a gap of 0.20 percentage points, smaller than the per-iteration standard deviation of the ensemble run---supporting the claim that competitive prompts can be discovered without exclusive reliance on large cloud models.

Table~\ref{tab:stats_summary} presents per-run descriptive statistics. The high standard deviation in all evolved runs ($\geq 1.4$ percentage points) reflects the stochastic nature of both LLM-driven
mutation and PassLLM inference. Notably, the ensemble run maintains a mean cracked rate of
$6.89\%$ -- already $3.4\times$ the baseline -- throughout the full 100-iteration run,
indicating that even non-peak prompts are substantially superior to the original.

\begin{table*}[t]
  \centering
  \footnotesize
  \caption{Descriptive statistics for each experimental run (excluding iteration 0).
    $n$: number of evaluated prompts; Mean, SD: mean and standard deviation of per-iteration
    cracked rate; Best CR: peak cracked rate; $\Delta$CR: improvement over baseline.}
  \label{tab:stats_summary}
  \begin{tabular}{lrrrrrr}
    \toprule
    \textbf{Run} & $n$ & \textbf{Mean CR} & \textbf{SD} & \textbf{Min CR} &
      \textbf{Best CR} & $\boldsymbol{\Delta}$\textbf{CR} \\
    \midrule
    Baseline         & --  & 2.02\% & --    & --    & 2.02\% &  --   \\
    Exp.~1 (Ollama)  & 100  & 4.13\% & 3.51\% & 0.00\% & 8.28\% & $+$4.1$\times$ \\
    Exp.~2 (Gemini)  & 100  & 3.79\% & 3.15\% & 0.00\% & 7.58\% & $+$3.75$\times$ \\
    Exp.~3 (Ensemble)& 100 & 6.89\% & 1.48\% & 1.07\% & \textbf{8.48\%} & $+$4.2$\times$ \\
    \bottomrule
  \end{tabular}
\end{table*}


Per-iteration CRs for all three runs alongside the baseline are shown in Fig.~\ref{fig:evolution_first45}. The ensemble achieves $\mathrm{CR} \geq 8\%$ at iteration 1, confirming that a single
LLM-driven mutation can already produce a substantial improvement. The Ollama runs also reach
8.28\% within 10 iterations budget, demonstrating remarkable efficiency.


The complete 100 iteration trajectory for the ensemble experiment in given in  the Fig.~\ref{fig:evolution_full}. The archive best (running maximum) plateaus after iteration 45, consistent with population convergence following the migration event at iteration 34.
The per-iteration trace shown high variance throughout with $\sigma = 1.48\%$, reflecting the
stochastic nature of LLM mutation. However, iterations above 8\% remain frequent as 30 out of 101
iterations achieve $\mathrm{CR} \geq 8\%$, thus confirming that the archive reliably captures strong prompts.

\subsection{Per-symbol F-score analysis}
\label{sec:fscore_results}

\begin{figure}[h]
\centering
\pgfplotsset{width=\linewidth, height=6cm}
\begin{tikzpicture}
  \begin{axis}[
    xlabel={Threshold $\tau$},
    ylabel={F-score},
    xmin=0, xmax=1,
    ymin=0, ymax=0.22,
    grid=major,
    legend pos=north east,
    legend style={font=\footnotesize},
    title={Per-symbol F-score vs.\ threshold},
    y tick label style={/pgf/number format/fixed, /pgf/number format/precision=3},
    scaled y ticks=false,
  ]
    \addplot[dashed, gray, thick, mark=*, mark size=1pt] coordinates {
      (0,0.022637)(0.05,0.099326)(0.10,0.115760)(0.15,0.122880)(0.20,0.117305)
      (0.25,0.110304)(0.30,0.100607)(0.35,0.087938)(0.40,0.076312)(0.45,0.066953)
      (0.50,0.057039)(0.55,0.049553)(0.60,0.040762)(0.65,0.033791)(0.70,0.030038)
      (0.75,0.022967)(0.80,0.018829)(0.85,0.010151)(0.90,0.005741)(0.95,0.002561)
    };
    \addplot[orange, mark=square*, mark size=1pt] coordinates {
      (0,0.022637)(0.05,0.163424)(0.10,0.179396)(0.15,0.164764)(0.20,0.145209)
      (0.25,0.125883)(0.30,0.105691)(0.35,0.086679)(0.40,0.072289)(0.45,0.058734)
      (0.50,0.047897)(0.55,0.038802)(0.60,0.029739)(0.65,0.025177)(0.70,0.020557)
      (0.75,0.013986)(0.80,0.009253)(0.85,0.006393)(0.90,0.002883)(0.95,0.001282)
    };
    \addplot[teal, mark=triangle*, mark size=1pt] coordinates {
      (0,0.022881)(0.05,0.155641)(0.10,0.191164)(0.15,0.182982)(0.20,0.162410)
      (0.25,0.143618)(0.30,0.119973)(0.35,0.107197)(0.40,0.084195)(0.45,0.070124)
      (0.50,0.058340)(0.55,0.049342)(0.60,0.040698)(0.65,0.033035)(0.70,0.024895)
      (0.75,0.019405)(0.80,0.014775)(0.85,0.011663)(0.90,0.006951)(0.95,0.003484)
    };
    \legend{Baseline, Exp.~2 (Gemini), Exp.~3 (Ensemble)}
  \end{axis}
\end{tikzpicture}
\caption{Per-symbol F-score curves for the baseline and best evolved prompts.
  Higher AUC indicates better alignment of the generated password's character-frequency
  distribution with that of real passwords. The ensemble prompt achieves the best AUC (0.074),
  while the baseline (0.059) has a flat profile indicating poor character-level coverage.}
\label{fig:fscore}
\end{figure}
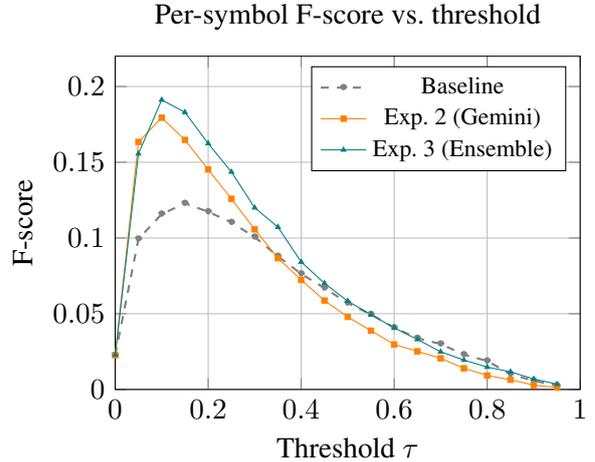

\noindent\textbf{To answer RQ4 (distribution realism),} we evaluate the best prompt from each configuration using the per-symbol F-score curves and AUC defined in Section~\ref{sec:fscore}. Fig.~\ref{fig:fscore} shows the per-symbol F-score curves (Eq.~\ref{eq:fscore}) for the best
prompts of the baseline, Experiment~2 (Gemini), and Experiment~3 (Ensemble) evaluated at thresholds
$\tau \in [0, 0.95]$. Table~\ref{tab:fscore_summary} summarizes the AUC values and peak scores.

\begin{table*}[t]
  \centering
  \caption{Per-symbol F-score performance. Peak values of the F-score, optimal threshold and AUC.}
  \label{tab:fscore_summary}
  \begin{tabular}{lccc}
    \toprule
    \textbf{Configuration} & \textbf{Peak F-score} & \textbf{Opt.\ $\tau$} & \textbf{AUC-F} \\
    \midrule
    Baseline          & 0.1229 & 0.15 & 0.0589 \\
    Exp.~2 (Gemini)   & 0.1794 & 0.10 & 0.0654 \\
    Exp.~3 (Ensemble) & \textbf{0.1912} & 0.10 & \textbf{0.0745} \\
    \midrule
    \multicolumn{2}{l}{Relative gain (Baseline $\to$ Ensemble):} & & $+$26.3\% \\
    \bottomrule
  \end{tabular}
\end{table*}

The ensemble prompt achieves a peak F-score of 0.191 at $\tau = 0.10$ -- a 65\% improvement over
the baseline peak of 0.123 (at $\tau = 0.15$). Two observations stand out. First, the baseline's
lower optimal threshold ($\tau = 0.15$) indicates that the original prompt produces passwords with
a coarser frequency distribution, only matching the most dominant characters. Second, both evolved
prompts peak at $\tau = 0.10$, suggesting that they better represent mid-frequency characters
(those appearing moderately in real passwords), which are precisely the characters that
discriminate real passwords from naive guesses.

The AUC-F improvement from 0.059 (baseline) to 0.074 (ensemble) represents a 26\% gain, confirming
that the higher cracked rate of evolved prompts is not merely due to over-generating highly common
passwords, but reflects a genuine improvement in character-distribution coverage.
\section{Discussion}
\label{sec:discussion}

Experimental results firstly underline the improvement rapidity. The ensemble achieves
$\mathrm{CR} \geq 8\%$ already at first iteration, confirming that even a single LLM-backed
mutation of the original prompt produces qualitatively better instructions. This 
gain reflects the sub-optimal nature of the baseline, which provides limited task-specific guidance by design. Methodologically, the baseline contradicts the conclusions of the work in~\cite{zou2025password} where prompt content has minimal impact on guessing performance. We attribute the baseline limitations primarily to \textit{manual} search of prompts with similar specificity levels, whereas the proposed evolution procedure enables explorations beyond the baseline setup.

The choice of MAP-Elites feature dimensions proves consequential. Substituting
\texttt{prompt\_length} for \texttt{diversity} in Experiment 3 yielded the best result,
suggesting that effective prompts occupy a characteristic length range specific enough to
conduct \textit{attack strategies}, yet concise enough to remain syntactically coherent, and that this
range is captured more faithfully by a length-based grid than by a diversity-based one.
Convergence behavior further support the value of the islands model. The archive-best improved
from 8.27\% to 8.48\% immediately after the migration, consistent with
the cross-island exchange. After
more iterations the archive gains stability and shows that 100 iterations are insufficient to
escape the resulting basin. Longer runs or a larger island counts ($K \geq 5$) may push the bound even further.

The near-parity between the local \texttt{qwen3:8b} with 8.28\% and the cloud ensemble
with 8.48\% is practically significant. The 0.20 point gap is acceptable within one
standard deviation of the ensemble run at a $\sigma = 1.48\%$. This confirms that the method
is viable in self-hosted, air-gapped deployments, which is an important consideration for
organizations with privacy constraints like banks or healthcare institutions. From an LLM-safety perspective, evolutionary prompt optimisation parallels evidence that context manipulation and short textual interventions can substantially redirect model outputs~\cite{afonin2026emergent,zou2023universal}, motivating hazard-aware release practices for auditing tooling~\cite{shevlane2023extremerisks}. Scope, threats to validity, and ethical constraints are summarised in Section~\ref{sec:limitations}. Future work includes testing on additional password distributions, extending to PII-based and password-reuse scenarios~\cite{zou2025password}, exploring richer island constructions, and investigating transfer of evolved prompts across corpora without re-optimisation. A defensive inversion is to employ the same framework to evolve password policies that minimise attacker CR, reinforcing existing password security protocols.

\section{Limitations and Ethics}
\label{sec:limitations}

\subsection{Limitations}
This study is a proof of concept that contains multiple constraints.

\begin{itemize}
  \item \textbf{Attack scenario.} In this evaluation, the focus is on the trawling setting from PassLLM~\cite{zou2025password}. It does not include targeted modes where a person uses PII to guess a password. If future work occurs, researchers can test if the optimization of prompts through evolution creates a gain that stays the same during targeted configurations.
  \item \textbf{Evolutionary stack and models.} There is a reliance on OpenEvolve~\cite{openevolve2025} for all evolutionary runs while the set of mutator and evaluator models is constant. As researchers use alternative quality-diversity algorithms, the outcomes are likely to change. By substituting other instruction-tuned models for the local mutators, the performance will also vary.
  \item \textbf{Generalisation and robustness.} For this data, the results come from a single RockYou-derived split. There is a high variance in the CR for each iteration because the mutations from the LLM happen by chance. On a single seed the study shows 100 iterations---but it is not clear how often the same high rates occur when a person uses other corpora.
  \item \textbf{Stable gains.} To show that the improvements are reliable, a person needs to use broader hyperparameter grids. For the MAP-Elites, the feature maps are also limited. Due to the current computational budgets, the stability of those gains is a topic for future work.
\end{itemize}

\subsection{Ethics}
We study behaviours that could lower the cost of offline guessing if misused. Experiments use only public leaked-password research artefacts (RockYou split) and do not involve live accounts. The contribution is methodological: we do not publish raw evolved prompts or turnkey guess lists, so reproducing peak numbers requires re-running the pipeline with non-trivial compute and domain setup. The upfront warning on the title page flags LLM-generated harmful text that may appear in mutation logs. We align dual-use reporting with norms in password research~\cite{florencio2014password}: informing defenders about attacker capability trajectories outweighs withholding aggregate evidence, provided release minimises operational enablement.

\section{Conclusion}
\label{sec:conclusion}

We proposed a methodology for automated evolutionary inference prompts optimization for
the LLM password guessing task through integration of the PassLLM core with the OpenEvolve
 framework. The proposed threat model situates the adversary among offline password-guessing scenarios discussed in the password-research literature~\cite{florencio2014password} and among LLM-supported security workflows surveyed more broadly~\cite{zhang2025llmcybersecurity}.

Experiments with three configurations of locally inferred 8B models, cloud models and ensembles
 show increase from 3.75 to 4.20$\times$ improvements in CR surpassing the baseline (2.02\%
to up to 8.48\%). With the local-only setup the setup achieves 8.28\% in as few as up to 10 iterations.
Additional per-symbol F-scoring confirms that evolved prompts not only found more correct passwords, they produced character frequency distributions that are substantially close to real-world
scenarios as experiments show with a 26\% increase in terms of AUC.

Taken together, the evidence in Section~\ref{sec:results} answers RQ1--RQ4 from Section~\ref{sec:rqs} affirmatively within the scope and caveats of Section~\ref{sec:limitations}.

Therefore, prompt evolving frameworks matter more than assumed as the original PassLLM prompts are showing sub-optimal performance, while evolution procedures rapidly
discover better approaches. 

From a defensive standpoint, adaptive password policies embedded into authentication systems alongside mandatory secondary factors complement ongoing efforts to characterise LLM misuse and safety trade-offs~\cite{shi2024llmsafety} and remain a practical line of defence as attack pipelines gain automated optimisation~\cite{shevlane2023extremerisks}.

\bibliography{custom}

\appendix
\section{Per-iteration cracked rate}
\label{sec:appendix}

\begin{figure}[h]
\centering
\pgfplotsset{width=\linewidth, height=6.1cm}
\begin{tikzpicture}
  \begin{axis}[
    xlabel={Iteration},
    ylabel={Cracked rate (\%)},
    xmin=0, xmax=45,
    ymin=0, ymax=15,
    grid=major,
    legend pos=north east,
    legend style={font=\footnotesize},
    unbounded coords=jump,
    y tick label style={/pgf/number format/fixed, /pgf/number format/precision=1},
    scaled y ticks=false,
  ]
    \addplot[dashed, gray, thick]
      table[x=iteration, y=baseline, col sep=comma] {pic/evolution_cracked_rate.csv};
    \addplot[blue, mark=*, mark size=1.2pt]
      table[x=iteration, y=Ollama_8.28, col sep=comma] {pic/evolution_cracked_rate.csv};
    \addplot[orange, mark=square*, mark size=1.2pt]
      table[x=iteration, y=Gemini_7.58, col sep=comma] {pic/evolution_cracked_rate.csv};
    \addplot[teal, mark=triangle*, mark size=1.2pt]
      table[x=iteration, y=Ensemble_8.48, col sep=comma] {pic/evolution_cracked_rate.csv};
    \legend{Baseline, Exp.~1 (Ollama), Exp.~2 (Gemini), Exp.~3 (Ensemble)}
  \end{axis}
\end{tikzpicture}
\caption{Per-iteration cracked rate (first 45 iterations). Baseline: PassLLM with the original
  prompt (2.02\%). All three evolved runs rapidly exceed the baseline; the ensemble first attains
  8.48\% at iteration 45.}
\label{fig:evolution_first45}
\end{figure}
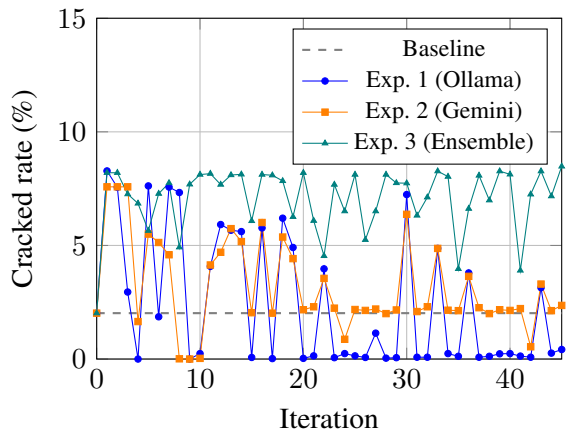

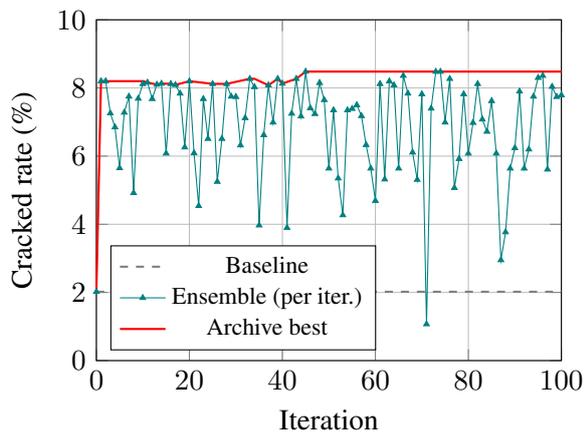
\begin{figure}[h]
\centering
\pgfplotsset{width=\linewidth, height=6.1cm}
\begin{tikzpicture}
  \begin{axis}[
    xlabel={Iteration},
    ylabel={Cracked rate (\%)},
    xmin=0, xmax=100,
    ymin=0, ymax=10,
    grid=major,
    legend pos=south west,
    legend style={font=\footnotesize},
    unbounded coords=jump,
    y tick label style={/pgf/number format/fixed, /pgf/number format/precision=1},
    scaled y ticks=false,
  ]
    \addplot[dashed, gray, thick]
      table[x=iteration, y=baseline, col sep=comma] {pic/evolution_cracked_rate.csv};
    \addplot[teal, mark=triangle*, mark size=1.2pt]
      table[x=iteration, y=Ensemble_8.48, col sep=comma] {pic/evolution_cracked_rate.csv};
    \addplot[red, thick, no marks] coordinates {
      (0,2.02)(1,8.2)(10,8.2)(11,8.16)(13,8.1)(14,8.13)(16,8.12)(17,8.09)
      (20,8.2)(25,8.12)(28,8.12)(33,8.27)(34,8.27)(37,8.08)(39,8.27)(40,8.13)
      (43,8.27)(45,8.48)(100,8.48)
    };
    \legend{Baseline, Ensemble (per iter.), Archive best}
  \end{axis}
\end{tikzpicture}
\caption{Full 100-iteration ensemble run. Teal: per-iteration cracked rate; red: running archive
  best (best CR seen up to each iteration). Migrations occurred at iterations 10, 34, 64,
  and 94; the migration at iteration 34 is associated with the step from 8.27\% to 8.48\%.}
\label{fig:evolution_full}
\end{figure}

This appendix presents detailed per-iteration cracked rates for the three evolved configurations 
(Ollama, Gemini, Ensemble) and the baseline across the first 45 iterations (Figure~\ref{fig:evolution_first45}), 
as well as the full 100-iteration ensemble run (Figure~\ref{fig:evolution_full}). 

As shown, the maximum cracked rate of 8.48\% is achieved by the ensemble at iteration~45, 
after which the archive best does not exceed this value, indicating convergence.

\end{document}